\documentclass[prd,reprint,showpacs,showkeys]{revtex4}
\usepackage{amsfonts}
\usepackage{mdframed}
\usepackage{amssymb}
\usepackage{amsmath}
\usepackage{graphicx}
\usepackage[font={footnotesize,it}]{caption}

\begin{document}

\title{Repeated crossing of two concentric spherical thin-shells with charge}
\author{S. Habib Mazharimousavi}
\email{habib.mazhari@emu.edu.tr}
\author{M. Halilsoy}
\email{mustafa.halilsoy@emu.edu.tr}
\affiliation{Department of Physics, Eastern Mediterranean University, Gazima\u{g}usa,
north Cyprus, Mersin 10, Turkey. }
\date{\today }

\begin{abstract}
Interaction / collision of two concentric spherical thin-shells of linear
fluid resulting in collapse has been considered recently. We show that
addition of finely tuned electric charges on the shells apart from the
cosmological constant serves to delay the collapse indefinitely, yielding an
ever colliding system of two concentric fluid shells. Given the finely tuned
charges this provides an example of a perpetual two-body motion in general
relativity.
\end{abstract}

\maketitle

\section{Introduction}

Due to frictional energy loss, perpetual motion in Newtonian mechanics
remains a myth, unless periodic pumping of the lost energy is supplemented
for. It remains still challenging to provide a prototype example of
perpetual motion in physics. A stable system of two (or more) objects may
emerge as a result of presumably attained perpetual motion. We direct our
attention next to general relativity in search for evidence which governs
the dynamics of heavenly objects; their stability, evolution, chaotic
behavior etc. Our two-body problem will be an exceptionally simple, yet
non-trivial one; two concentric spherical thin-shells each made of a gas
fluid. Even in such a simple system of concentric binary shells one
confronts naturally with repeated passing through / crossing of the shells.
This is a highly non-linear process that admittedly realistic solution can
hardly be found unless simplifying assumptions are imposed. The exact
two-body problem can be solved in Newtonian mechanics. The reason is the
consistent definition of a center of mass of the system and absence of
gravitational radiation. In general relativity on the other hand the
existence of gravitational radiation prevents us to define reliably center
of mass of objects in motion. We note that special system at rest are
available in general relativity; the charged masses along a line found by
Majumdar and Papapetrou is one such example \cite{1,2,3}. Interaction of two
(or more) objects otherwise turns into a highly inelastic process of
collision associated with radiation. In such a process the total energy of
interacting masses alongside with gravitational and other fields, such as
electromagnetic, scalar, spinor, axion etc. transmute gravity into the
formation of a newly curved spacetime. We have exact solutions of colliding
waves in general relativity that describes evolution / head-on collision of
two null fields \cite{4}. We recall that when two null planar shells (both
thin and thick) collide the result is a naked singularity. For such an exact
process see \cite{5} and references cited therein. When we focus our
attention to non-null fields moving slower than light speed we encounter
serious technical problems. Collision of shells, branes and bubbles in
general relativity have been considered before. Some relatively older works
are given in Ref. \cite{6,7,8,9,10,11,12,13,14,15,16} while some more recent
studies are given in Ref. \cite{17,18,19,20,21,22,23,24,25,26,27,28,29,30,31}%
. In addition to the collision between two shells, Wang and Gao have studied
recently the collision between a static spherically symmetric thin-shell
wormhole and a thin shell \cite{32}. In their study they have applied the
formalism given by Langlois, et al on the conservation laws for collisions
of branes and shells in general relativity \cite{33}.

In this paper we restrict ourselves to a relatively simple problem of
interacting / colliding / crossing of two spherical shells both made of a
linear gas. Such a gas / fluid is described by the equation of state (EoS) $%
P=w\sigma $ in which $P$ and $\sigma $ are the pressure and the energy
density of the shells, respectively, and $w$ is a constant. The energy
conditions (weak, strong and dominant) are satisfied for $-\frac{1}{3}<w<1$.
The recent interest in such a problem of spherical shell-crossings provides
the main motivation for the present study \cite{30,31}. The process, as
advocated by those authors will be a rather special one, namely, the four
velocity of each shell remains unchanged during the collision process. Yet
the mass / energy (and electric charge added in this paper) of each shell
will effect the other shell to create a new spacetime in analogy with the
colliding null fields. For the energy conservation and junction conditions
we appeal to EoS and the Israel conditions \cite{34,35,36,37,38}. The
thin-shell of gas taking part in a collision process is said to satisfy
transparency condition provided that only gravitational interaction is
active during the crossing.

The conclusion drawn with the addition of charge to the shells is that
oppositely charged shells tend to form a perpetually colliding pair at
almost equal distances so that the system points toward a non-collapsing
one. Iteration of the exact solution beyond certain number of crossings
becomes technically out of our reach. Yet the numerical analysis to certain
orders of collisions provides convincing evidence that the pattern tends
with finely tuned charges to a non-collapsing system of shells crossing each
other indefinitely. It is observed that the positive effect played by the
negative cosmological constant in \cite{30,31} is extended further by the
presence of opposite charges on the shells.

Organization of the paper is as follows. In Section II we introduce our
formalism. The transparency of collision is described in Section III. The
effect of charge is introduced in Section IV. Our paper ends with Conclusion
in Section V.

\section{The Formalism}

Let's start with two concentric timelike shells identified as $\Sigma _{i}$
(inner shell) and $\Sigma _{o}$ (outer shell) within a spherically symmetric
bulk spacetime. These shells divide the bulk into three different regions
given by $\mathcal{M}_{1}$ (inside the inner shell)$,$ $\mathcal{M}_{2}$
(between the two shells) and $\mathcal{M}_{3}$ (outside the outer shell)
each expressed with the following line element%
\begin{equation}
ds_{a}^{2}=-f_{a}\left( r_{a}\right) dt_{a}^{2}+\frac{dr_{a}^{2}}{%
f_{a}\left( r_{a}\right) }+r_{a}^{2}d\Omega _{a}^{2}
\end{equation}%
in which $a=1,2,3$ and $d\Omega _{a}^{2}=d\theta _{a}^{2}+\sin ^{2}\theta
_{a}d\phi _{a}^{2}.$ Also the induced line element on the shells can be
written as 
\begin{equation}
ds_{i}^{2}=-d\tau _{i}^{2}+r_{i}^{2}d\Omega _{i}^{2}
\end{equation}%
and 
\begin{equation}
ds_{o}^{2}=-d\tau _{o}^{2}+r_{o}^{2}d\Omega _{o}^{2}
\end{equation}%
in which $\tau _{i/o}$ is the proper time on each shell and $r_{i/o}$ is the
location of each shell. Note that the two shells are specified with
different coordinates. In the vicinity of inner shell one finds $\Sigma
_{i}=\Sigma _{i}^{\left( 1\right) }=\Sigma _{i}^{\left( 2\right) }$, $%
r_{1}=r_{2}=r_{i},$ $\theta _{1}=\theta _{2}=\theta _{i},$ $\phi _{1}=\phi
_{2}=\phi _{i}$ while $t_{1}\neq t_{2}$ and we identify 
\begin{equation}
\left. -f_{1}\left( r_{1}\right) dt_{1}^{2}+\frac{dr_{1}^{2}}{f_{1}\left(
r_{1}\right) }\right\vert _{\Sigma _{i}^{\left( 1\right) }}=\left.
-f_{2}\left( r_{2}\right) dt_{2}^{2}+\frac{dr_{2}^{2}}{f_{2}\left(
r_{2}\right) }\right\vert _{\Sigma _{i}^{\left( 2\right) }}=-d\tau _{i}^{2}
\end{equation}%
which in turn implies that%
\begin{equation}
\left( \frac{dt_{1,2}}{d\tau _{i}}\right) ^{2}=\frac{f_{1,2}\left(
r_{i}\right) +\left( \frac{dr_{i}}{d\tau _{i}}\right) ^{2}}{%
f_{1,2}^{2}\left( r_{i}\right) }.
\end{equation}%
The normal unit four vectors to the inner shell on either side (from $%
\mathcal{M}_{1}$ toward $\mathcal{M}_{2}$) are given by%
\begin{equation}
\left( n_{\mu }^{\left( 1,2\right) }\right) _{i}=\frac{1}{\sqrt{\Delta
^{\left( 1,2\right) }}}\left( -\frac{dr_{i}/d\tau _{i}}{dt_{1,2}/d\tau _{i}}%
,1,0,0\right) ,
\end{equation}%
where%
\begin{equation}
\Delta ^{\left( 1,2\right) }=\frac{f_{1,2}^{2}\left( r_{i}\right) }{%
f_{1,2}\left( r_{i}\right) +\left( \frac{dr_{i}}{d\tau _{i}}\right) ^{2}}.
\end{equation}%
The second fundamental form or the extrinsic curvature tensor of the inner
shell is obtained with the nonzero components given by%
\begin{equation}
\left( K_{\tau }^{\tau \left( 1,2\right) }\right) _{i}=\frac{2\frac{%
d^{2}r_{i}}{d\tau _{i}^{2}}+f_{1,2}^{\prime }}{2\sqrt{f_{1,2}\left(
r_{i}\right) +\left( \frac{dr_{i}}{d\tau _{i}}\right) ^{2}}},
\end{equation}%
\begin{equation}
\left( K_{\theta }^{\theta \left( 1,2\right) }\right) _{i}=\left( K_{\phi
}^{\phi \left( 1,2\right) }\right) _{i}=\frac{\sqrt{f_{1,2}\left(
r_{i}\right) +\left( \frac{dr_{i}}{d\tau _{i}}\right) ^{2}}}{r_{i}}
\end{equation}%
in which a 'prime' denotes $\frac{d}{dr}.$ From the Israel junction
conditions \cite{34,35,36,37,38} 
\begin{equation}
\left[ K_{A}^{B}\right] -\left[ K\right] \delta _{A}^{B}=-8\pi GS_{A}^{B},
\end{equation}%
with $\left[ Z\right] =Z^{\left( 2\right) }-Z^{\left( 1\right) }$, $\left[ K%
\right] =trace\left[ K_{A}^{B}\right] $ and $S_{A\left( i\right)
}^{B}=diag\left( -\sigma _{i},P_{i},P_{i}\right) $ which is the energy
momentum tensor on the inner shell, one finds%
\begin{equation}
\sigma _{i}=\frac{1}{4\pi G}\left( \frac{\sqrt{f_{1}\left( r_{i}\right)
+\left( \frac{dr_{i}}{d\tau _{i}}\right) ^{2}}}{r_{i}}-\frac{\sqrt{%
f_{2}\left( r_{i}\right) +\left( \frac{dr_{i}}{d\tau _{i}}\right) ^{2}}}{%
r_{i}}\right) ,
\end{equation}%
\begin{equation}
P_{i}=\frac{1}{8\pi G}\left( \frac{2\frac{d^{2}r_{i}}{d\tau _{i}^{2}}%
+f_{2}^{\prime }}{2\sqrt{f_{2}\left( r_{i}\right) +\left( \frac{dr_{i}}{%
d\tau _{i}}\right) ^{2}}}-\frac{2\frac{d^{2}r_{i}}{d\tau _{i}^{2}}%
+f_{1}^{\prime }}{2\sqrt{f_{1}\left( r_{i}\right) +\left( \frac{dr_{i}}{%
d\tau _{i}}\right) ^{2}}}+\frac{\sqrt{f_{2}\left( r_{i}\right) +\left( \frac{%
dr_{i}}{d\tau _{i}}\right) ^{2}}}{r_{i}}-\frac{\sqrt{f_{1}\left(
r_{i}\right) +\left( \frac{dr_{i}}{d\tau _{i}}\right) ^{2}}}{r_{i}}\right) .
\end{equation}%
Note also that, the energy conservation for the inner shell demands that 
\begin{equation}
\frac{d\sigma _{i}}{dr_{i}}=-\frac{2}{r_{i}}\left( P_{i}+\sigma _{i}\right) .
\end{equation}%
A similar analysis for the outer shell yields%
\begin{equation}
\left( \frac{dt_{2,3}}{d\tau _{o}}\right) ^{2}=\frac{f_{2,3}\left(
r_{o}\right) +\left( \frac{dr_{o}}{d\tau _{o}}\right) ^{2}}{%
f_{2,3}^{2}\left( r_{o}\right) },
\end{equation}%
\begin{equation}
\sigma _{o}=\frac{1}{4\pi G}\left( \frac{\sqrt{f_{2}\left( r_{o}\right)
+\left( \frac{dr_{o}}{d\tau _{o}}\right) ^{2}}}{r_{o}}-\frac{\sqrt{%
f_{3}\left( r_{o}\right) +\left( \frac{dr_{o}}{d\tau _{o}}\right) ^{2}}}{%
r_{o}}\right) ,
\end{equation}%
\begin{equation}
P_{o}=\frac{1}{8\pi G}\left( \frac{2\frac{d^{2}r_{o}}{d\tau _{o}^{2}}%
+f_{3}^{\prime }}{2\sqrt{f_{3}\left( r_{o}\right) +\left( \frac{dr_{o}}{%
d\tau _{o}}\right) ^{2}}}-\frac{2\frac{d^{2}r_{o}}{d\tau _{o}^{2}}%
+f_{2}^{\prime }}{2\sqrt{f_{2}\left( r_{o}\right) +\left( \frac{dr_{o}}{%
d\tau _{o}}\right) ^{2}}}+\frac{\sqrt{f_{3}\left( r_{o}\right) +\left( \frac{%
dr_{o}}{d\tau _{o}}\right) ^{2}}}{r_{o}}-\frac{\sqrt{f_{2}\left(
r_{o}\right) +\left( \frac{dr_{o}}{d\tau _{o}}\right) ^{2}}}{r_{o}}\right) ,
\end{equation}%
and 
\begin{equation}
\frac{d\sigma _{o}}{dr_{o}}=-\frac{2}{r_{o}}\left( P_{o}+\sigma _{o}\right)
\end{equation}%
in which the label $o$ stands for the outer shell and $r_{o}$ is the
location of the outer shell.

Next, we consider the EoS on each shell to be that of a linear gas i.e., 
\begin{equation}
P_{i/o}=w_{i/o}\sigma _{i/o}
\end{equation}%
in which $w_{i/o}$ is a constant. Considering the EoS (18) with the energy
conservation (13) and (17) one finds%
\begin{equation}
\sigma _{i/o}=\frac{C}{r_{i/o}^{2\left( w_{i/o}+1\right) }}
\end{equation}%
in which $C$ is an integration constant. To compare our numerical results
with \cite{30,31} we set $C=\frac{m_{o/i}l^{2w_{i/o}}}{4\pi G}$ in which $%
m_{o/i}$ is a new constant and $\frac{1}{l^{2}}=-3\Lambda $ in which $%
\Lambda $ is the cosmological constant to appear in our metric functions.
For inner shell one finds the equation of motion, using (11), as%
\begin{equation}
\left( \frac{dr_{i}}{d\tau _{i}}\right) ^{2}+V_{i}=0
\end{equation}%
with 
\begin{equation}
V_{i}=\frac{f_{2}+f_{1}}{2}-\frac{\left( f_{2}-f_{1}\right) ^{2}}{4\nu
_{i}^{2}}-\frac{\nu _{i}^{2}}{4}
\end{equation}%
in which%
\begin{equation}
\nu _{i}=4\pi Gr_{i}\sigma _{i}=\frac{m_{i}l^{2w_{i}}}{r_{i}^{2w_{i}+1}}
\end{equation}%
where $\sigma _{i}$ is given in (19). A similar equation can be found for
the outer shell such that 
\begin{equation}
\left( \frac{dr_{o}}{d\tau _{o}}\right) ^{2}+V_{o}=0
\end{equation}%
in which%
\begin{equation}
V_{o}=\frac{f_{3}+f_{2}}{2}-\frac{\left( f_{3}-f_{2}\right) ^{2}}{4\nu
_{o}^{2}}-\frac{\nu _{o}^{2}}{4}
\end{equation}%
with 
\begin{equation}
\nu _{o}=4\pi Gr_{o}\sigma _{o}=\frac{m_{o}l^{2w_{o}}}{r_{o}^{2w_{o}+1}}.
\end{equation}%
The time evolution of the two thin shells are with respect to two different
proper times i.e., $\tau _{o}$ and $\tau _{i}$ and therefore we can not
study their collision in terms of a common time. Therefore we use the
relations (5) and (20)/(23) to write%
\begin{equation}
\frac{dr_{i}}{d\tau _{i}}=\frac{dr_{i}}{dt_{2}}\sqrt{\frac{f_{2}\left(
r_{i}\right) -V_{i}}{f_{2}^{2}\left( r_{i}\right) }}
\end{equation}%
and%
\begin{equation}
\frac{dr_{o}}{d\tau _{o}}=\frac{dr_{o}}{dt_{2}}\sqrt{\frac{f_{2}\left(
r_{o}\right) -V_{o}}{f_{2}^{2}\left( r_{o}\right) }}.
\end{equation}%
Finally we plug these into (20) and (23) to determine the evolution of the
thin-shells in terms of the common time $t_{2}$ given by%
\begin{equation}
\left( \frac{dr_{i/o}}{dt_{2}}\right) ^{2}+\frac{f_{2}^{2}\left(
r_{i/o}\right) V_{i/o}}{f_{2}\left( r_{i/o}\right) -V_{i/o}}=0.
\end{equation}%
%
%
%
%
%
%
%
%
%
%
%
%
%
%
%
%
%
%%%%%%%%%%%%%%%%%%%%%%%%%%%%%%%%%%%%%%%%%%%%%%%%%%%%%%%%%%%%%%%%%%%%%%%%%%%%%
\begin{figure}[tbp]
\includegraphics[width=100mm,scale=0.7]{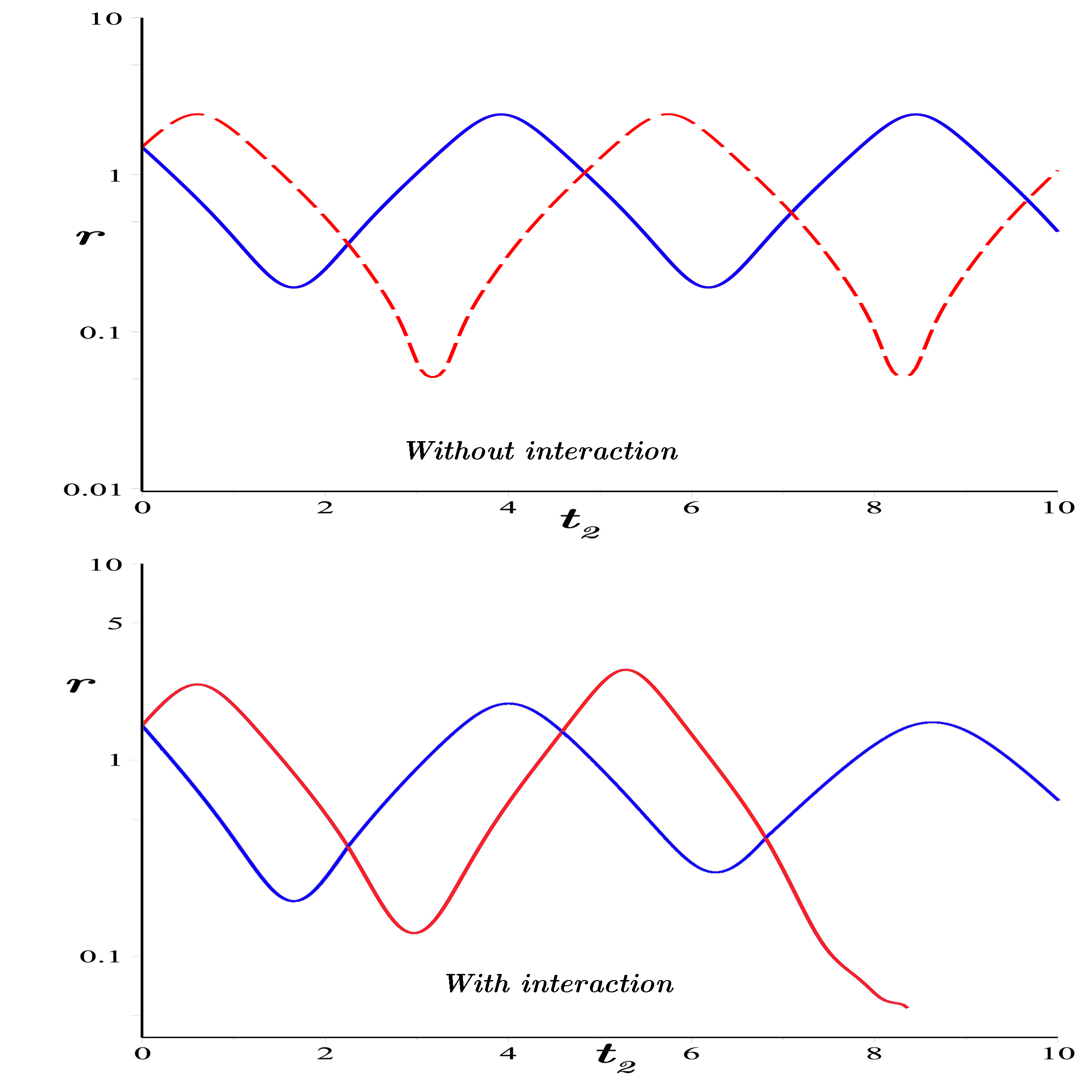}
\caption{Time evolution of each shell independently are depicted at the top
figure. Time evolution of the inner and outer shells when both shells are
chargeless but interact gravitationally are shown at the bottom figure. We
choose $Q_{a}=0,$ $M_{1}=0.00,$ $M_{2}=0.025$, $M_{3}=0.05$, $%
m_{o/i}=0.0136, $ $w_{o/i}=0.2$, $l=1$ and the initial position of both
shells set to be $r_{o/i}\left( t_{2}=0\right) =1.50$. The vertical axis is
logarithmic.}
\end{figure}
%%%%%%%%%%%%%%%%%%%%%%%%%%%%%%%%%%%%%%%%%%%%%%%%%%%%%%%%%%%%%%%%%%%%%%%%%%%%%%%%%
\begin{figure}[tbp]
\includegraphics[width=100mm,scale=0.7]{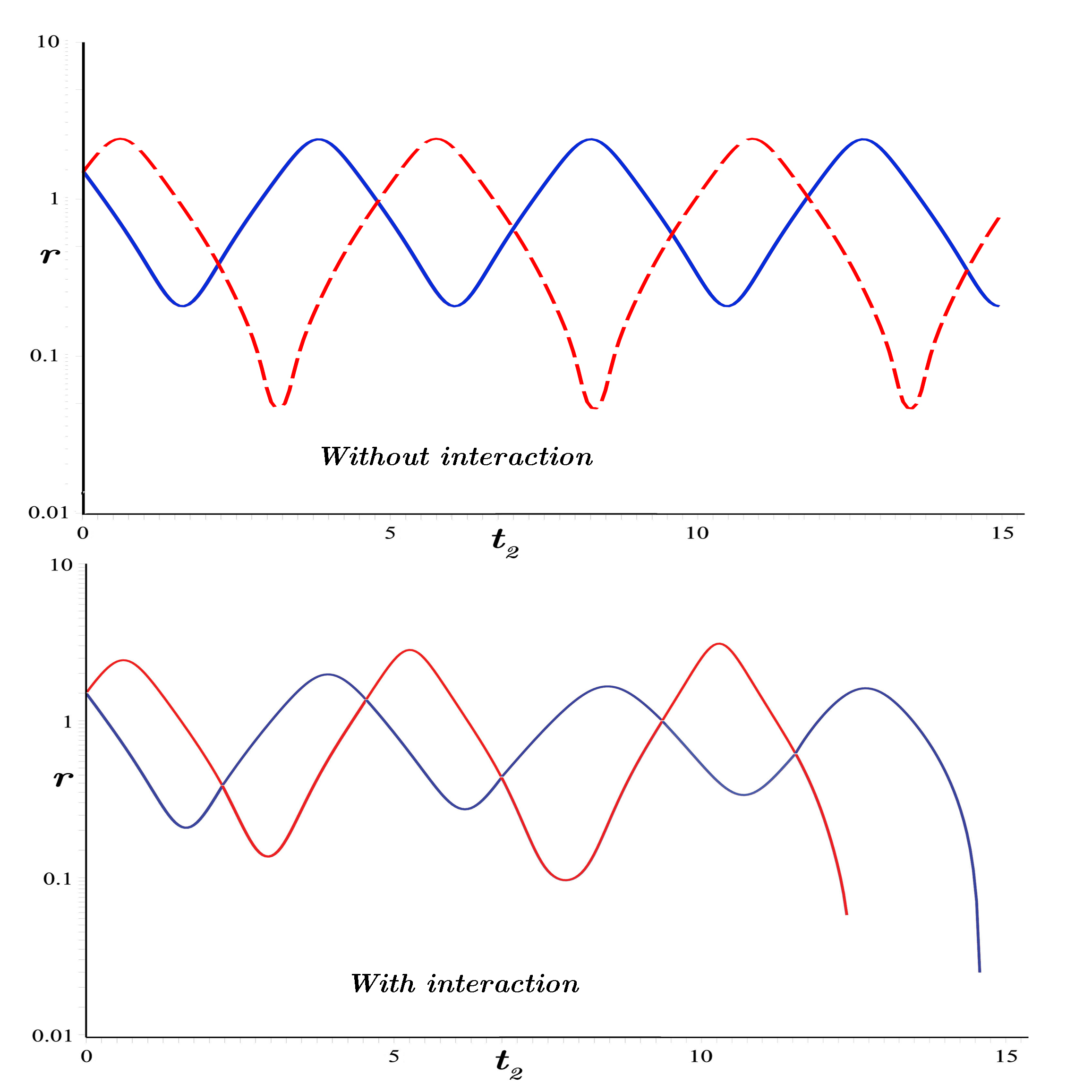}
\caption{Time evolution of each shell independently are depicted are shown
at the top figure. Time evolution of the inner and outer shells when one
shell (initially inner) is charged while the other is chargeless are shown
at the bottom figure. We have $Q_{1}=0$, $Q_{2}=Q_{3}=0.15,$ $M_{1}=0.00,$ $%
M_{2}=0.025$, $M_{3}=0.05$, $m_{o/i}=0.0136,$ $w_{o/i}=0.2$, $l=1$ and the
initial position of both shells set to be $r_{o/i}\left( t_{2}=0\right)
=1.50 $. The vertical axis is logarithmic.}
\end{figure}
%%%%%%%%%%%%%%%%%%%%%%%%%%%%%%%%%%%%%%%%%%%%%%%%%%%%%%%%%%%%%%%%%%%%%%%%%%%%%%%%%%%%
\begin{figure}[tbp]
\includegraphics[width=100mm,scale=0.7]{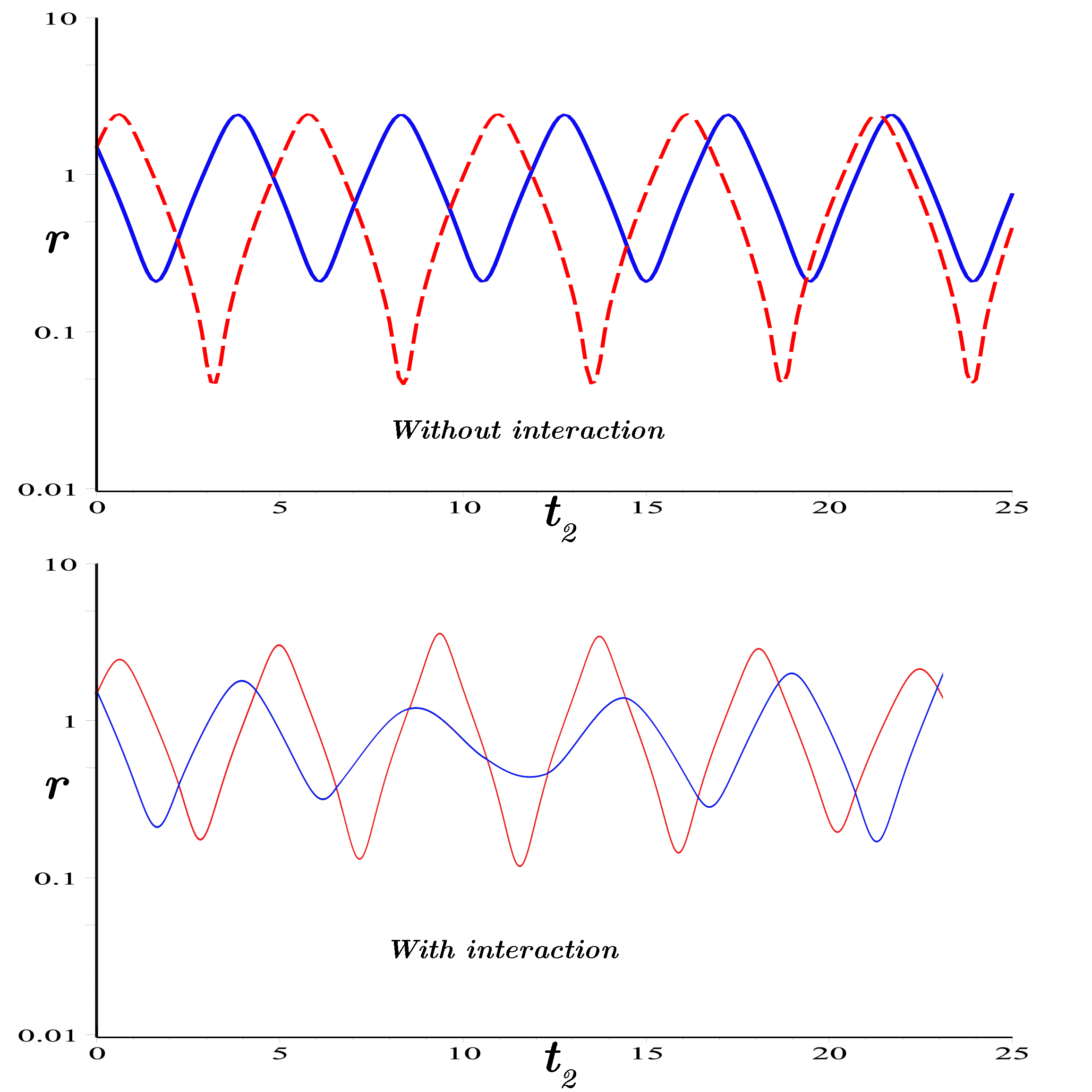}
\caption{Time evolution of each shell independently are depicted at the top.
Time evolution of the inner and outer shells when both shells are charged
are given at the bottom figure. Precisely, $Q_{1}=0$, $Q_{2}=-Q_{3}=-0.15,$ $%
M_{1}=0.00,$ $M_{2}=0.025$, $M_{3}=0.05$, $m_{o/i}=0.0136,$ $w_{o/i}=0.2$, $%
l=1$ and the initial position of both shells set to be $r_{o/i}\left(
t_{2}=0\right) =1.50$. The vertical axis is logarithmic.}
\end{figure}
%%%%%%%%%%%%%%%%%%%%%%%%%%%%%%%%%%%%%%%%%%%%%%%%%%%%%%%%%%%%%%%%%%%%%%%%%%%

\section{Transparency of the collision process}

In this section we consider a transparent collision between the two shells
which has been studied by Ida and Nakao in \cite{39}. According to the
condition of transparency, the four velocities of the shells are conserved
during the collision i.e. the four velocities are continuous. Without going
through the detail of the calculation we follow \cite{39} and consider the
two shells passing through each other and divide the bulk spacetime into
three parts which are labeled as $\mathcal{M}_{1},$ $\mathcal{M}_{4\text{ }}$%
and $\mathcal{M}_{3}$ such that the inner shell after the collision was the
outer shell before the collision and vice versa. Therefore the spacetime on
the sides of $\Sigma _{o/i}$ are given by (1) but $a=1,4$ for $\Sigma _{o}$
and $a=4,3$ for $\Sigma _{i}.$ Applying the transparent collision conditions
one finds the line element of $\mathcal{M}_{4}$ given by (1) with \cite{39} 
\begin{multline}
f_{4}\left( r\right) =\frac{\nu _{i}^{2}+\nu _{o}^{2}}{2}+\frac{%
f_{1}+f_{3}-f_{2}}{2}+\frac{\left( \nu _{i}^{2}-f_{1}\right) \left( \nu
_{o}^{2}-f_{3}\right) }{2f_{2}}+ \\
\frac{\epsilon _{i}\epsilon _{o}}{2f_{2}}\sqrt{\nu _{i}^{4}-2\left(
f_{1}+f_{2}\right) \nu _{i}^{2}+\left( f_{1}-f_{2}\right) ^{2}}\sqrt{\nu
_{o}^{4}-2\left( f_{2}+f_{3}\right) \nu _{o}^{2}+\left( f_{2}-f_{3}\right)
^{2}}
\end{multline}%
in which $\epsilon _{i/o}=\pm 1$ stands for the direction of the motion of
the shells, i.e., if the shell is moving radially outward /inward $\epsilon
_{i/o}=+1/-1.$ Let's add that after the collision the equation found in
previous sections are all valid provided the substitutions $%
o\longleftrightarrow i$, and $2\rightarrow 4$ are made in all equations. For
instance, the equation of motion of the shells after the collision is given
by%
\begin{equation}
\left( \frac{dr_{o/i}}{dt_{2}}\right) ^{2}+\frac{f_{4}^{2}\left(
r_{o/i}\right) V_{o/i}}{f_{4}\left( r_{o/i}\right) -V_{o/i}}=0
\end{equation}%
in which%
\begin{equation}
V_{o/i}=\frac{f_{4}+f_{1/3}}{2}-\frac{\left( f_{4}-f_{1/3}\right) ^{2}}{4\nu
_{o/i}^{2}}-\frac{\nu _{o/i}^{2}}{4}
\end{equation}%
with%
\begin{equation}
\nu _{o/i}=4\pi Gr_{o/i}\sigma _{o/i}=\frac{m_{o/i}l^{2w_{o/i}}}{%
r_{o/i}^{2w_{o/i}+1}}.
\end{equation}%
Let's add that, in Eq. (30) a sub $o$ implies that the shell was initially
the outer shell. After the first collision, although we still use the same
label physically the outer shell becomes the new inner shell. Therefore
following to each collision the order of the shells changes and no matter
what we set them initially (which is indicated by their sub-indices $o$ or $%
i $) they can be inner or outer shells with time. Finally we would like to
add that the above formalism is applicable for a generic spherically
symmetric bulk spacetime with the line element given in (1). Therefore in
addition to the vacuum spherically symmetric bulk one may also consider the
case with energy momentum tensor which due to spherical symmetry reads as%
\begin{equation}
T_{\mu }^{\nu }=diag\left( -T,-T,\tilde{T},\tilde{T}\right)
\end{equation}%
in which $T$ and $\tilde{T}$ are found by applying the Einstein equations in
separate regions.

\section{Interacting charged shells}

As stated in Introduction, our aim is to investigate the effect of electric
/ magnetic charge on the collapsing shells introduced by \cite{30,31}. To
accomplish this we consider the spacetimes to be Reissner-Nordstr\"{o}m AdS
with a line element given by (1) while%
\begin{equation*}
f_{a}=1-\frac{2M_{a}}{r}+\frac{Q_{a}}{r^{2}}+\frac{r^{2}}{l^{2}}
\end{equation*}%
in which $a=1,2,3$ refer to the inside, in between and outside of the
shells, respectively, before each collision. After each collision the metric
function of inside/$a=1$ and outside /$a=3$ remain unaltered but the metric
between the shells changes as given by (29) and this sequence either goes on
forever or ends with a collapse. In Fig.1 we reproduced the case reported in 
\cite{30,31} with $Q_{a}=0,$ $M_{1}=0.00,$ $M_{2}=0.025$, $M_{3}=0.05$, $%
m_{o/i}=0.0136,$ $w_{o/i}=0.2$, $l=1$ and initial position of both shells
are set to be at $r_{o/i}\left( t_{2}=0\right) =1.50.$ As it was reported in 
\cite{30,31} the two shells after making three collisions eventually
collapse and a black hole is formed. In Fig. 2 we solve numerically the same
two-shells system with charges on one of the shells only. We choose $Q_{1}=0$%
, $Q_{2}=Q_{3}=0.15$ and the other parameters including the initial
conditions are the same as in Fig. 1. The charges on the shells can be
calculated by using Gauss's law which yields the charge on the inner and
outer shells to be $q_{i}=0.15$ and $q_{o}=0$ initially. We observe that
adding positive charge on one of the shells and keeping the rest of
configuration the same causes the system collide two more times but
eventually collapsing to make a charged black hole as the former case. In
our Fig. 3 we have added charges on both shells. Initially the charge on
inner and outer shells are chosen to be $q_{i}=-0.015$ and $q_{o}=0.030$
which in turn implies $Q_{1}=0$, $Q_{2}=-Q_{3}=-0.015$. The other parameters
and initial positions are as in Figs. 1 and 2. We observe that the shells
initially collide quite the same as the other two cases but owing to the
opposite charges on the shells the collapse does not occur and the shells
after making finite number of crossings return almost to their initial
conditions. This is what we may interpret as a stable configuration which
repeats itself periodically but eternally. Let's end this section by adding
that at the top of each figure, for comparison, we also present the motion
of each shell independent of the other.

We would like to add that, in this specific study our aim was to show that a
collapsing system of repeated crossing of two concentric chargeless
spherical thin-shells may be stabilized without altering the masses of the
shells but adding enough opposite charges on the shells. As the nature of
the collisions and the equations are highly non-linear we could not find an
analytic correlations between the masses and charges necessary but as we
have shown numerically in the specific examples, such charges exist.
Correlation between masses and added charges can only be obtained by more
number of collisions and tabular account for both quantities.

\section{Conclusion}

The dynamics of two concentric spherical binary thin-shells made of gas
satisfying a linear equation of state is revisited. As a new element in this
paper we added electric charge to the shells and investigated their
evolution / collision in a suitable time variable. The problem was
considered previously without charge but with a negative cosmological
constants in Ref. \cite{30,31}. The confining boundary role of AdS in \cite%
{30,31} is shown to be valid also in the presence of finely-tuned opposite
charges on the shells. Adding charge on one shell did not change the
ultimate collapse of the binary-shell system but given the finely tuned
charges on both shells with opposite signs we obtain a non-collapsing,
permanently colliding system. Such a system may have impact in cosmological
formations as an example of self-sustaining two-body problem with perpetual
motion.

\end{document}